\documentclass[runningheads]{llncs}
\usepackage{graphicx}
\usepackage[frozencache,cachedir=.]{minted}
\usepackage{xspace}
\usepackage{bussproofs}
\usepackage{wrapfig}
\usepackage{amssymb}
\usepackage{hyperref}
\newcommand{\code}[1]{\texttt{#1}}
\newcommand{\hasteeplus}{HasTEE$^+$\xspace} 

\begin{document}
\title{\hasteeplus: Confidential Cloud Computing and Analytics with Haskell}
\author{Abhiroop Sarkar\orcidID{0000-0002-8991-9472} \and
Alejandro Russo\orcidID{0000-0002-4338-6316}}
\authorrunning{Abhiroop Sarkar, Alejandro Russo}
\institute{
Chalmers University, Gothenburg, Sweden\\
\email{\{sarkara,russo\}@chalmers.se}}
\maketitle              % typeset the header of the contribution
\begin{abstract}
% Context
Confidential computing is a security paradigm that enables the protection of confidential code and data in a co-tenanted cloud deployment using specialized hardware isolation units called Trusted Execution Environments (TEEs).
By integrating TEEs with a Remote Attestation protocol, confidential computing allows a third party to establish the integrity of an \textit{enclave} hosted within an untrusted cloud.
% Gap
However, TEE solutions, such as Intel SGX and ARM TrustZone, offer low-level C/C++-based toolchains that are susceptible to inherent memory safety vulnerabilities and lack language constructs to monitor explicit and implicit information-flow leaks.
Moreover, the toolchains involve complex multi-project hierarchies and the deployment of hand-written attestation protocols for verifying \textit{enclave} integrity.

% Innovation
\hspace{\parindent} We address the above with \hasteeplus, a domain-specific language (DSL) embedded in Haskell that enables programming TEEs in a high-level language with strong type-safety.
\hasteeplus assists in multi-tier cloud application development by (1) introducing a \textit{tierless} programming model for expressing distributed client-server interactions as a single program, (2) integrating a general remote-attestation architecture that removes the necessity to write application-specific cross-cutting attestation code, and (3) employing a dynamic information flow control mechanism to prevent explicit as well as implicit data leaks.
We demonstrate the practicality of \hasteeplus through a case study on confidential data analytics, presenting a data-sharing pattern applicable to mutually distrustful participants and providing overall performance metrics.

\end{abstract}
\section{Introduction} \label{sec:intro}
Confidential computing \cite{DBLP:conf/seed/MulliganPSSV21} is an emerging security paradigm that ensures the isolation of sensitive computations and data during processing, shielding them from potential threats within the underlying infrastructure. 
It accomplishes this by employing specialised hardware isolation units known as Trusted Execution Environments (TEEs). 
A TEE unit, such as the Intel SGX \cite{DBLP:conf/isca/McKeenABRSSS13} or ARM TrustZone \cite{armtz}, provides a \textit{disjoint} region of code and data memory that allows for the physical isolation of a program's execution and state from the underlying operating system, hypervisor, I/O peripherals, BIOS and other firmware.

TEEs further allow a remote party to establish \textit{trust} by providing a measurement of the sensitive code and data, composing a signed \textit{attestation report} that can be verified.
As such, TEEs have been heralded as a leading contender to enforce a strong notion of \textit{trust} within cloud computing infrastructure\cite{DBLP:journals/tocs/BaumannPH15,DBLP:journals/cacm/RussinovichCFCD21,DBLP:journals/accs/ZegzhdaUNP17}, particularly in regulated industries such as healthcare, law and finance \cite{DBLP:journals/fcomp/GeppertDSE22}.

However, an obstacle in the wide-scale adoption of confidential computing has been the awkward programming model of TEEs \cite{sgxdep}. 
TEEs, such as Intel SGX, involve partitioning the state of the program into a trusted project linked with Intel-supplied restricted C standard library \cite{tlibc} and an untrusted project that communicates with the trusted project using custom data copying protocols \cite{sgxcopying}. 
The complexity is compounded for distributed, multi-tiered cloud applications due to the semantic friction in adhering to various data formats and protocols across multiple projects \cite{DBLP:journals/corr/abs-2207-08019}, resulting in increased developer effort \cite{northwood2018full}. 

Furthermore, a well-known class of security vulnerabilities \cite{cowan2000buffer,DBLP:conf/ccs/Shacham07} arise from the memory-unsafety of the TEE projects. 
Given the security-critical nature of TEE applications, efforts have been made to introduce Rust-based \cite{DBLP:conf/ccs/WangWDSJDLZWL19} and Golang-based SDKs \cite{DBLP:conf/usenix/GhosnLB19} aimed at mitigating memory unsafety vulnerabilities.
However, the same applications remain vulnerable to unintended information leaks \cite{DBLP:journals/jsac/SabelfeldM03}. Consider the following \textit{trusted} function hosted within a TEE:

\begin{minted}[fontsize=\footnotesize, linenos]{C}
#include "library.h" //provides `int computeIdx(char *)`
char *secret = "...";
int public_arr [15] = ....;
void trusted_func(char *userinput, int inputsize) {
   if (memcmp(secret, userinput, inputsize) == 0) {
     int val = computeIdx(userinput);
     ocall_printf("%d\n", public_arr[val]); // handwritten printf routine
   } else { ocall_printf("0\n"); }
}
\end{minted}

In the above program, at least five attack vectors are present - (1) the \code{inputsize} parameter can be abused to cause a buffer-overflow attack, (2) the \code{userinput} parameter can be tampered by a malicious operating system to force an incorrect branching, (3) even with the correct \code{inputsize} and \code{userInput}, the attacker can observe \code{stdout} to learn which branch was taken, (4) the trusted library function - \code{computeIdx} - could intentionally (if written by a malicious party) or accidentally leak \code{secret}, and (5) finally the attacker can use timing-based side channels to learn the branching information or even the \code{secret} \cite{DBLP:journals/ieeesp/ChenCXZLL20}.

Our contribution through this work is -- \textit{\hasteeplus} -- a Domain Specific Language (DSL) embedded in Haskell and targeted towards confidential computing. 
\hasteeplus is designed to mitigate at least four of the five attack vectors mentioned above. Additionally, it offers a \textit{tierless} programming model, thereby simplifying the development of distributed, multi-tiered cloud applications.

\hasteeplus builds on the HasTEE \cite{DBLP:conf/haskell/SarkarKRC23} project, which crucially provides a \textit{Glasgow Haskell Compiler (GHC)} runtime \cite{DBLP:conf/icfp/MarlowJS09} capable of running Haskell on Intel SGX machines. 
While using a memory-safe language like Haskell or Rust mitigates attack vector (1), all the other attack vectors remain open in the HasTEE project.
\hasteeplus notably adds support for a general one-time-effort remote-attestation infrastructure that helps mitigate the attack vector (2).
While the underlying protocol employs Intel's RA-TLS \cite{knauth2018integrating}, \hasteeplus ensures that programmers are not required to create custom attestation code for capturing and sending measurements or conducting integrity checks \cite{sgxraexample}.

For attack vectors (3) and (4), \hasteeplus adds support for a dynamic information flow control (IFC) mechanism based on the Labeled IO Monad (LIO) \cite{DBLP:conf/haskell/StefanRMM11}. 
Accordingly, we adopt a \textit{floating-label} approach from OSes such as HiStar \cite{DBLP:conf/osdi/ZeldovichBKM06}, enabling \hasteeplus to relax some of the impractical I/O restrictions in the original HasTEE project.
Side-channel attacks (attack vector (5)) remain outside the scope of \hasteeplus.

Concerning the complex programming model, modern TEE incarnations like AMD SEV-SNP \cite{sev2020strengthening} and Intel TDX \cite{inteltdx} introduce a virtualization-based solution, opting to virtualize the entire project instead of partitioning it into trusted and untrusted components. 
At the cost of an increased trusted code base, this approach simplifies the TEE project layout. Nevertheless, it remains vulnerable to the complexities of a multi-tiered cloud application, as well as all five of the aforementioned attack vectors.

The \textit{tierless} programming model of \hasteeplus expresses multi-client-server projects as a single program and uses the Haskell type system to distinguish individual clients. 
A separate monadic type, such as \code{Client "client1" a}, demarcates each individual client, while HasTEE's multi-compilation tactic partitions the program.
This programming model, evaluated on Intel SGX, remains applicable on newer Intel TDX machines.

\textbf{Contributions.} We summarize the key contributions of \hasteeplus here:

\begin{itemize}
    \item \hasteeplus introduces a \textit{tierless} DSL (Section~\ref{tierless}), capable of expressing multi-tiered confidential computing applications as a single program, increasing an application's comprehensibility and reducing developer effort.

    \item \hasteeplus incorporates a remote attestation design (Section~\ref{attestation}) that relieves programmers from crafting custom integrity checks and attestation setups.
    
    \item \hasteeplus integrates dynamic information flow control mechanisms (Section \ref{ifc}) to prevent explicit and implicit information leaks from applications.

    \item We use \hasteeplus's IFC mechanism and cryptography in a \textit{data clean room} case study (Section~\ref{dcr}), showing a general data sharing pattern for conducting analytics on confidential data among mutually distrustful participants.
\end{itemize}

\section{Threat Model}\label{sec:threat}

We build upon the threat model of the HasTEE project \cite{DBLP:conf/haskell/SarkarKRC23} and other related works on Intel SGX \cite{DBLP:conf/osdi/ArnautovTGKMPLM16,DBLP:journals/tocs/BaumannPH15,DBLP:conf/usenix/GhosnLB19,DBLP:conf/ccs/WangWDSJDLZWL19}. 
In such a threat model, an attacker attempts to compromise the code and data memory within the TEE. 
The attacker has administrative access to the operating system, hypervisor and other related system software hosted on a malicious cloud service.

We expand the above threat model to include an \textit{active attacker} attempting to compromise the integrity of the data flowing into the TEE, as well as a \textit{passive attacker} who observes the public channels that the trusted software interacts with to learn more about its behaviour. 
The threat model terminology is adopted from the $J_E$ project \cite{DBLP:conf/csfw/OakABS21}, and related attacks are discussed in subsequent sections. 
Another class of potential threats emerge from the inclusion of public software libraries into TEE software, such as cryptography libraries, which might accidentally or intentionally leak secrets \cite{cryptocve}.

\hasteeplus's attestation infrastructure, based on Intel's RA-TLS protocol \cite{knauth2018integrating} accounts for masquerading attacks \cite{goldman2006linking,stumpf2006robust}.
Availability attacks such as denial-of-service and hardware side-channel attacks are outside the scope of this work.

\section{The \hasteeplus DSL}

We illustrate the key APIs of \hasteeplus using a password checker application in Listing~\ref{lst:pwdchecker}, and explain the individual types and functions in the subsequent sections. 
Notably, the entire application, consisting of a separate client and server, can be expressed within 27 lines of Haskell code (excluding import declarations).

%% A paragraph on monads - Mary, Ale

\begin{listing}[!ht]
\begin{minted}[fontsize=\footnotesize, linenos]{haskell}
pwdChecker :: EnclaveDC (DCLabeled String) -> String -> EnclaveDC Bool
pwdChecker pwd guess = do
  l_pwd <- pwd
  priv  <- getPrivilege
  p     <- unlabelP priv l_pwd
  if p == guess then return True else return False

data API = API { checkpwd :: Secure (String -> EnclaveDC Bool) }

client :: API -> Client "client" ()
client api = do
  liftIO $ putStrLn "Enter your password:"
  userInput <- liftIO getLine
  res <- gatewayRA ((checkpwd api) <@> userInput)
  liftIO $ putStrLn ("Login returned " ++ show res)

app :: App Done
app = do
  pwd   <- inEnclaveLabeledConstant pwdLabel "password"
  let priv = toCNF "Alice"
  efunc <- inEnclave (dcDefaultState priv) $ pwdChecker pwd
  runClient (client (API efunc))
  where
    pwdLabel :: DCLabel
    pwdLabel = "Alice" %% "Alice" :: DCLabel

main = runAppRA "client" app >> return ()  
\end{minted}
\caption{A password checker application in \hasteeplus}
\label{lst:pwdchecker}
\end{listing}

Listing~\ref{lst:pwdchecker} shows the user \textit{Alice} storing her password in the TEE memory and deploying the trusted function \code{pwdChecker} to conduct a password check.
%
% The application uses three notable monads - \code{EnclaveDC}, \code{Client} and \code{App}, as in HasTEE \cite{DBLP:conf/haskell/SarkarKRC23}.
The application uses three key types -- \code{EnclaveDC}, \code{Client}, and \code{App}, adapted from HasTEE \cite{DBLP:conf/haskell/SarkarKRC23}. All three types implements a monadic interface, denoted as \code{m}, constructed using fundamental operations \code{return :: a -> m a} and \code{(>>=) :: m a -> (a -> m b) -> m b} (read as \textit{bind}). The \code{return x} operation produces a computation returning the value of x without side effects, while the \code{(>>=)} function sequences computations and their associated side effects. In Haskell, we often use the \textit{do-notation} to express such monadic computations.

The \code{EnclaveDC} monad represents the trusted computations that get loaded onto a TEE. 
The name \textit{Enclave} alludes to an Intel SGX \textit{enclave}, while \textit{DC} stands for \textit{Disjunction Category}, which we further explain in Section~\ref{ifc}.

The client-side of the application is represented by the namesake \code{Client} monad that captures a \textit{type-level} string - \code{"client"}. 
The type-level string allows the Haskell type-system to distinguish between multiple clients and serves as an identifier for the monadic computation runner function - \code{runAppRA} on line 27.

The monad \code{App} serves as a staging area where the enclave data (\code{"password"}) and the enclave computation (\code{pwdChecker}) are loaded into trusted memory. 
Also, the \code{client} function is provided with the \code{API} it will use to communicate with the TEE within the \code{App} monad.
We provide the type signatures of a simplified subset of the \hasteeplus APIs used in Listing~\ref{lst:pwdchecker} for loading trusted data and computations, and running the computations, in Fig.~\ref{fig:loadapi} below.

\begin{figure}[!ht]
\centering
\begin{minted}[frame=single, fontsize=\footnotesize]{haskell}
inEnclave :: Label l => LIOState l p -> a -> App (Secure a)
inEnclaveLabeledConstant 
          :: Label l => l -> a -> App (EnclaveDC (DCLabeled a))
gatewayRA :: (Binary a, Label l)
          => Secure (Enclave l p a) -> Client loc a
(<@>)     :: Binary a => Secure (a -> b) -> a -> Secure b
runClient :: Client loc a -> App Done
runAppRA  :: Identifier   -> App a -> IO a
\end{minted}
\caption{\hasteeplus APIs for loading data and computations on the TEE and invoking the TEE (parameterized types simplified and typeclass constraints omitted for brevity).}
\label{fig:loadapi}
\end{figure}

The application begins with the \code{app} function (lines~17-25).
The functions \code{inEnclaveLabeledConstant} and \code{inEnclave} are used to load the trusted data (line~19) and the trusted function (line~21) in the enclave, respectively.
The \code{runClient} at line~22 runs the monadic \code{Client} computations. 
Note, there is no equivalent \code{runEnclave}, as in our programming model \textit{a client functions as the main application driver, while the enclave serves as a computational service.}

The client and server communicate with each other through a user-defined \code{API} type (line 8) that encapsulates a remote closure, represented using the \code{Secure} type constructor.
This closure is constructed on line 21 with the \code{inEnclave} function whose type signature can be found in Fig.~\ref{fig:loadapi}.
The parameter \code{LIOState l p} and the typeclass constraint \code{Label l} are explained in Section~\ref{ifc}.

The \code{client} function (lines 10-15) has access to the remote closure through the \code{API} type. 
The remote function is invoked on line~14 using the \code{<@>} operator to emulate function application and the \code{gatewayRA} function executes the remote function call. 
The respective type signatures are specified in Fig.~\ref{fig:loadapi}.

A notable type in the \code{pwdChecker} function is \code{DCLabeled String} that captures the password string but is labeled with ownership information of user \textit{Alice}. 
The labeling happens on line~19 using the \code{inEnclaveLabeledConstant} function and the label \code{pwdLabel} (lines~24,25). 
The body of \code{pwdChecker} uses certain IFC operations - \code{getPrivilege} and \code{unlabelP}, which we elaborate on in Section~\ref{ifc}.

The structure of Listing~\ref{lst:pwdchecker} represents the standard style of writing \hasteeplus programs, where the monadic types \code{EnclaveDC}, \code{Client} and \code{App} indicate the partitioning within the program body.
We discuss the partitioning tactic and the underlying representation to capture multiple clients in the next section.

\subsection{Tierless client-server programming}\label{tierless}

\begin{figure}[!ht]
         \centering
         \includegraphics[scale=0.18]{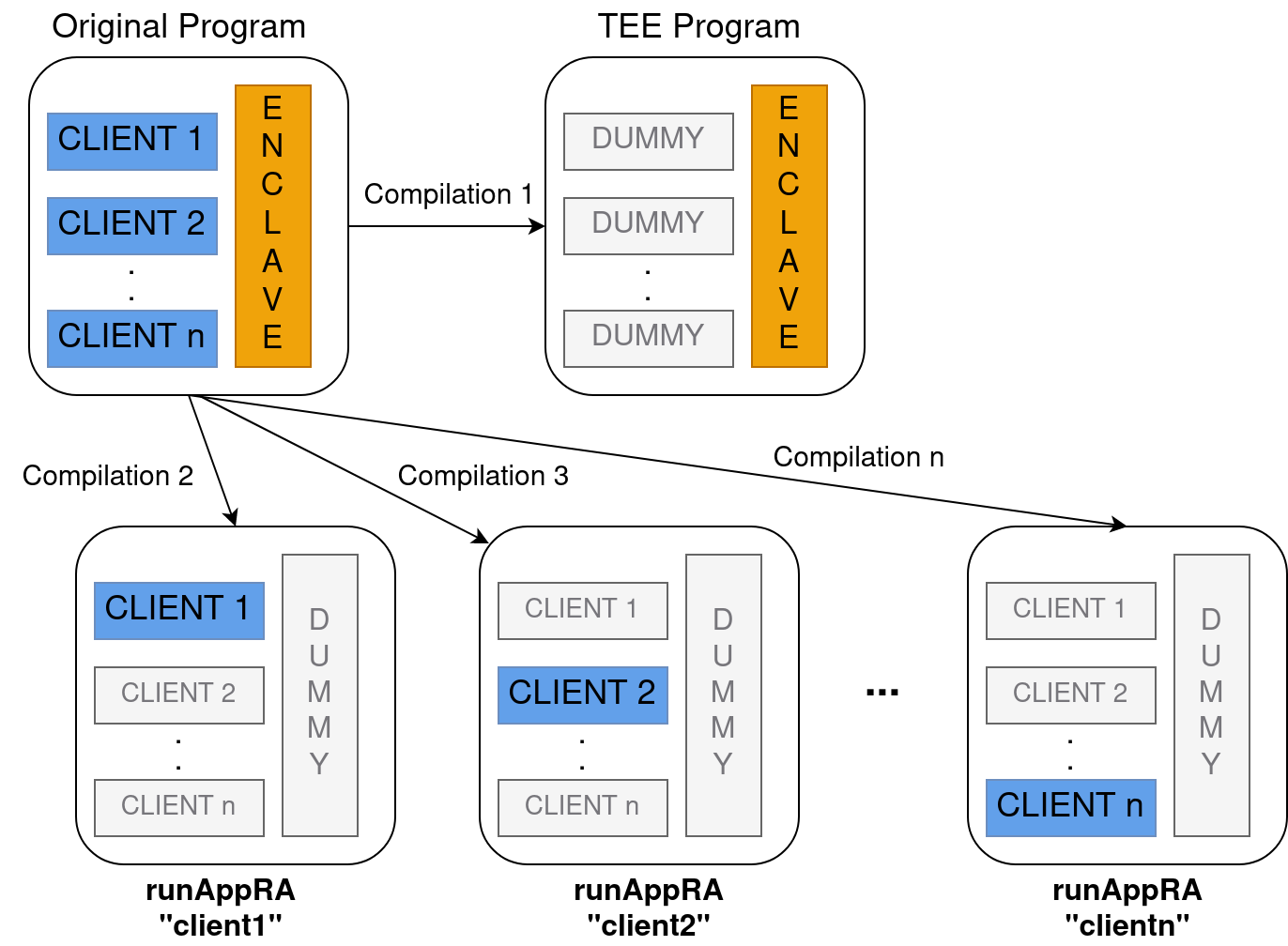}
         \caption{\hasteeplus's partitioning uses multiple compilations to create binaries that can dynamically dispatch the code for only one concerned monad based on a string identifier}
         \label{fig:partition}
\end{figure}

\hasteeplus builds on the partitioning strategy of HasTEE \cite{DBLP:conf/haskell/SarkarKRC23} but generalizes it for multiple clients. 
The strategy involves compiling the program \code{n} times for \code{n} parties.
The compilation of the enclave program substitutes dummy implementation for all client monads. Similarly, each client compilation substitutes a dummy value for the enclave monad. 
The distinction between each client is done using a type-level string identifier, such as \code{"client1"}. At runtime, this identifier is used to dynamically dispatch the correct client, as shown in Fig.~\ref{fig:partition}.

The dynamic identifier-based dispatch of the client computation is inspired from the HasChor library \cite{DBLP:journals/pacmpl/ShenKK23} for choreographic programming.
In this approach, a \code{Client} monad is parameterized with a type-level location string, which is used at runtime to execute the desired \code{Client} computation (see Listing~\ref{lst:client}).

\begin{listing}[!ht]
\begin{minted}[fontsize=\footnotesize]{Haskell}
data Client (loc :: Symbol) a where
  Client :: (KnownSymbol loc) => Proxy loc -> IO a -> Client loc a

symbolVal :: forall (n :: Symbol) proxy. KnownSymbol n => proxy n -> String

runClient :: Client loc a -> App Done
runClient (Client loc cl) = App $ do
  location <- get -- the underlying App monad captures the location
  if ((symbolVal loc) == location)
  then liftIO cl >> return Done
  else return Done -- cl not executed
\end{minted} 
\caption{The underlying Client monad in \hasteeplus}
\label{lst:client}
\end{listing}

The function \code{symbolVal} is provided by GHC to reflect types as terms at runtime, provided the types are constrained by the \code{KnownSymbol} typeclass.
The \code{runClient} function in Listing~\ref{lst:client} queries the App monad that captures the string within the underlying \code{App} monad.
The \code{runClient} implementation for the \code{EnclaveDC} module is simply implemented as \code{runClient \_ = return Done}, which amounts to a dummy implementation.
A case study involving multiple clients is demonstrated in Section~\ref{dcr}.

For the remote function invocation, the \code{inEnclave} function internally builds a \textit{dispatch table} mapping an integer identifier to each enclave function.
The client only gets access to the integer identifier. 
It uses the \code{<@>} operator to gather the function argument and the \code{gatewayRA} function to serialise the arguments, make the remote function call (specifying the identifier), and obtain the result computed on the remote enclave machine.
Note the \code{Binary} typeclass constraint (Fig.~\ref{fig:loadapi}) on both of the remote invocation functions for binary serialisation.

The complete implementation details of \hasteeplus has been made publicly available\footnote{\url{https://github.com/Abhiroop/HasTEE}}. 
Further details on the operational semantics of the general partitioning strategy can be found in the HasTEE\cite{DBLP:conf/haskell/SarkarKRC23} paper.

%%% ERASE FOR SPACE CRUNCH
% \noindent\textbf{Multiple Enclaves}. While our tierless DSL discusses multiple clients above, there appears to be a missing notion of \textit{multiple} enclaves. 
% %
% Multiple enclaves usually refer to multiple sets of encrypted memory pages hosted on - (1) the same virtual address space, (2) different virtual address spaces, and (3) different physical address spaces.

% Scenario (1) can be naturally expressed in \hasteeplus using a combination of the \code{forkOS} and \code{runInBoundThread} functions from GHC's \code{Control.Concurrent} library \cite{DBLP:conf/cefp/Marlow11}.
% %
% Regarding scenarios (2) and (3), we deliberately decided against representing disjoint address spaces as distinct types in our DSL. 
% %
% This is because those scenarios are simply considered a special case of the standard client-server interaction already expressible in \hasteeplus.

\subsection{Remote Attestation via a Monitoring Server} \label{attestation}

%%% FOR SPACE CRUNCH
%%% Straightaway start explaining RA-TLS but have to mention quotes somehwere

A key component of confidential computing is remote attestation, which establishes \textit{trust} on a TEE within a malicious environment. 
In \hasteeplus, we conduct our experiments on Intel SGX enclaves, and hence our infrastructure is integrated with the SGX attestation protocol. 
The low-level protocol is a multi-step process \cite{knauth2018integrating} that begins with the client sending a nonce to the TEE, the TEE then creates a manifest file that includes an ephemeral key to encrypt future communication. 
Next, the TEE  generates an \textit{attestation report} that summarizes the enclave and platform state. 
A quoting enclave on the same machine verifies and signs the report, now called a \textit{quote}, and returns it to the client. 
The client then communicates with the Intel Attestation Service (IAS) to verify the quote.

The API for this interface is quite low-level and involves programming at the level of a device driver (\code{/dev/attestation}).
Intel's RA-TLS protocol abstracts over the low-level APIs and presents an API deeply tied to the TLS protocol.
RA-TLS operates by extending an X.509 certificate to incorporate the attestation report within an unused X.509 extension field. 
During TLS connection setup, it uses the TLS handshake to transmit the \textit{quote}, calculated using the protocol described earlier \cite{knauth2018integrating}.
The enclave programmer, working with RA-TLS, interacts with a modified TLS implementation such as Mbed TLS \cite{mbedtls}. 
Now, the focus shifts back to dealing with low-level socket-programming APIs, such as:

%% ERASE FOR SPACE SQUEEZE
% \begin{figure}[!ht]
\begin{minted}[fontsize=\footnotesize]{C}
int (*ra_tls_create_key_and_crt_der_f)(uint8_t** der_key, size_t* der_key_size, 
                                       uint8_t** der_crt, size_t* der_crt_size);
void (*ra_tls_set_measurement_callback_f)(int (*f_cb)(const char* mrenclave, 
           const char* mrsigner, const char* isv_prod_id, const char* isv_svn));
\end{minted}
% \end{figure}

Once again, managing these APIs is \textit{error-prone} and \textit{memory unsafe}. 
Additionally, it requires constructing \textit{underspecified protocols}. 
Most importantly, the programmer is burdened with handling \textit{cross-cutting concerns} that are irrelevant to the application code.

In \hasteeplus, we abstract over Intel's RA-TLS protocol. 
As mentioned earlier,  clients always serve as the primary program \textit{driver}, while the enclave functions as a computational \textit{service}.
The \textit{enclave-as-a-service} model is implemented by representing the entire enclave program as an infinitely running server.
The server is implemented in C using the Mbed TLS library \cite{mbedtls}, which can parse and verify the modified X.509 certificate.
Internally, when the enclave runs, it spawns the C server hosted on the enclave memory but using separate memory pages. 
We use GHC's Foreign Function Interface to establish a communication channel between the C and Haskell heaps. 
Listing~\ref{lst:runappra} shows a high-level overview of the implementation.

\begin{listing}[!ht]
\begin{minted}[fontsize=\footnotesize]{Haskell}
runAppRA :: Identifier -> App a -> IO a
runAppRA ident (App s) = do
  (a, vTable) <- runStateT s (initAppState ident)
  flagptr  <- malloc :: IO (Ptr CInt)
  dataptr  <- mallocBytes dataPacketSize :: IO (Ptr CChar)
  _   <- forkOS (startmbedTLSSERVER_ffi tid flagptr dataptr)
  result <- try (loop vTable flagptr dataptr) -- exception handler
  -- exception handling and freeing C pointers
  return a
  where
    loop :: [(CallID, Method)] -> Ptr CInt -> Ptr CChar -> IO ()
    loop vTable flagptr dataptr = do -- body elided
      -- non-blocking loop that gets woken when data arrives;
      -- `flagptr` indicates data arrival; read data from `dataptr`
      -- invoke the correct method from the lookup table `vTable`
      loop vTable flagptr dataptr -- continue the event loop
-- implemented in C
startmbedTLSSERVER_ffi :: ThreadId -> Ptr CInt -> Ptr CChar -> IO ()
\end{minted} 
\caption{High-level template of the \code{runAppRA} function for the enclave}
\label{lst:runappra}
\end{listing}

\begin{figure}[!ht]
         \centering
         \includegraphics[scale=0.22]{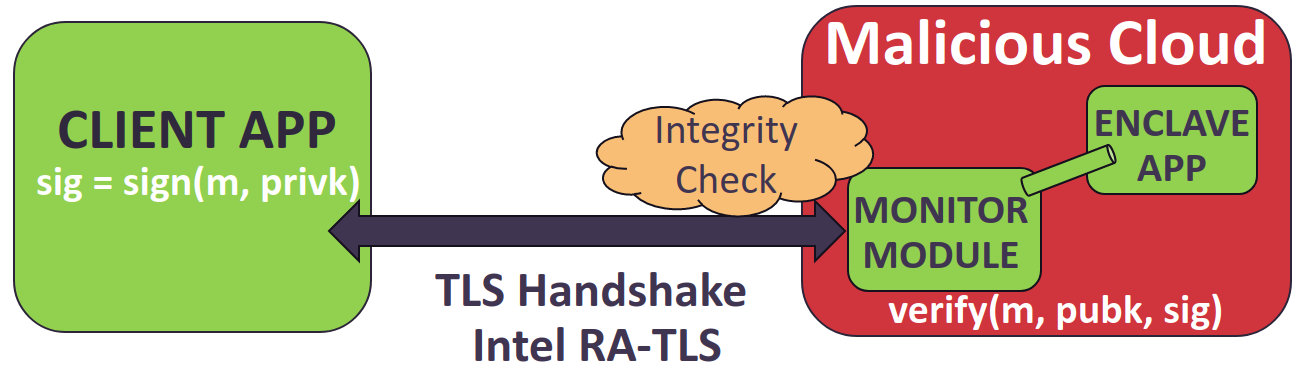}
         \caption{\hasteeplus's remote attestation infrastructure abstracts over Intel's RA-TLS protocol and supports establishing the identity of the client and the server}
         \label{fig:ra}
\end{figure}
% \vspace{-5mm}
The Mbed TLS-based C server module acts as a \textit{monitor} for the enclave application. 
All dataflows between the clients and the enclave pass through this module, which conducts integrity checks on incoming data at this point.
Fig.~\ref{fig:ra} shows the \hasteeplus general monitoring architecture.

There are two distinct attackers targeting the dataflow - (1) a malicious OS snooping or tampering with the data flowing into the enclave, and (2) a malicious client, potentially colluding with the OS, repeatedly sending garbage inputs to observe the behaviour of the enclave.
The TLS channel specifically prevents the first attack. 
For the second attack, we use public-key-cryptography-based digital signatures to verify the identity of the client making the request. 

For instance, in Listing~\ref{lst:pwdchecker}, we provision the public key for user \textit{Alice} during enclave boot time and disallow password checks from other malicious clients. 
While Intel-SGX also offers mutual attestation services, they depend on the client machine supporting an SGX enclave. 
Considering Intel's recent deprecation of SGX services on desktops and other client devices \cite{sgxdep}, our scheme aligns well.

\vspace{-3.5mm}
% In HasTEE, an integrity check is opt-in, depending on the performance implication, using the -fintegrity-check. 

\subsection{Dynamic Information Flow Control} \label{ifc}

While programming TEEs using a memory-safe and type-safe language inherently provides stronger guarantees than programming in C/C++, such TEE applications remain vulnerable to unintended information leaks.
To mitigate such explicit and implicit information leaks, \hasteeplus integrates a dynamic information flow control mechanism within the underlying \code{EnclaveDC} monad.
For instance, the trusted function \code{pwdChecker} from Listing~\ref{lst:pwdchecker} operates in this monad and captures a \textit{labeled} string - the user password. 
The internal representation of the types \code{EnclaveDC a} and \code{DCLabeled a} is as follows:

\begin{minted}[fontsize=\footnotesize]{Haskell}
newtype Label l => Enclave l p a = Enclave (IORef (LIOState l p) -> IO a)

data Labeled l t where
  LabeledTCB :: (Label l, Binary l, Binary t) => l -> t -> Labeled l t

type EnclaveDC = Enclave DCLabel DCPriv
type DCLabeled = Labeled DCLabel

data LIOState l p = LIOState { lioCurLabel :: l, lioClearance :: l
                             , lioOutLabel :: l, lioPrivilege :: Priv p}
\end{minted} 

The above representation is inspired by the LIO Haskell library \cite{DBLP:conf/haskell/StefanRMM11}. 
The \code{Enclave} monad is parameterized by the label type \code{l} and privilege type \code{p}, wherein the \code{Label} typeclass captures a lattice \cite{DBLP:journals/cacm/Denning76} with partial order $\sqsubseteq$ governing the allowed flows. 
The \code{Enclave} monad employs a \textit{floating label} approach, inspired by the HiStar OS \cite{DBLP:conf/osdi/ZeldovichBKM06}. 
In this approach, the \textit{computational context} retains a current label $L_{cur}$. 
Upon reading sensitive data labeled $L$, it \textit{taints} the context with the label $L_{cur} \sqcup L$, where $\sqcup$ denotes the least upper bound.
The floating label approach restricts subsequent effects and enforces the lattice property, thereby preventing information leaks from higher-classified data to lower contexts in the lattice—a principle known as non-interference \cite{DBLP:conf/sp/GoguenM82a}.
In the \code{LIOState} type, \code{lioCurLabel} represents $L_{cur}$, and \code{lioClearance} imposes an upper bound on the upward flow of a computational context within the lattice.

The \code{EnclaveDC} type specialises the \code{Enclave} monad to use \textit{disjunction category (DC)} labels \cite{DBLP:conf/nordsec/StefanRMM11}.
A DC Label captures both the \textit{confidentiality} \cite{DBLP:journals/cacm/Denning76}  and \textit{integrity} \cite{Biba} as a tuple.
It employs the notion of mutually distrusting \textit{principals}, whose conjunction represents restrictions on both the confidentiality and integrity of the data. 
An example label type is found in Listing~\ref{lst:pwdchecker}, where \code{pwdLabel} constructs a DC Label using the tuple-construction operator \code{\%\%} and a string representation of the principal \textit{Alice} to give \code{"Alice" \%\% "Alice"} (line~25).

A notable component in the DC Label system is the notion of a \textit{privilege}, which is the type parameter \code{p} in the \code{Enclave} monad.
In most real-world scenarios, the strict enforcement of non-interference is impractical, and privileges allow relaxing this policy by defining a more flexible ordering relation, $\sqsubseteq_P$.

\begin{wrapfigure}{r}{0.5\textwidth}
\begin{center}
\begin{prooftree}
\AxiomC{$P \land C_2 \Longrightarrow C_1$}
\AxiomC{$P \land I_1 \Longrightarrow I_2$}
\BinaryInfC{$<C_1, I_1>\ \sqsubseteq_P\ <C_2, I_2>$}
\end{prooftree}
\end{center}
\end{wrapfigure}

\noindent In \hasteeplus's DCLabel implementation, we use a \textit{conjunctive normal form (CNF)} representation for each disjunctive category of confidentiality and integrity. 
Hence, given a boolean formula P representing the privileges and two labels $<C_1, I_1>$ and $<C_2, I_2>$, the $\sqsubseteq_P$ is defined as shown in the formula.

In Listing~\ref{lst:pwdchecker}, we use the \code{toCNF} function (line~20) to generate a privilege for \textit{Alice} to declassify the password, provisioning it to the \code{Enclave} monad at boot time (line~21).
This allows the \code{pwdChecker} function to invoke \code{getPrivilege} (line~4) and then use that privilege to call the \code{unLabelP} function (line~5), which internally computes the $\sqsubseteq_P$ formula shown above.
The \code{unlabelP} function and a related set of core APIs for \hasteeplus's IFC enforcement is shown in Fig.~\ref{fig:difcapi}.
Implementation note: for prototyping, we represent principals and corresponding privileges using \code{String}s. In practice, a 512-bit private-key-hash is recommended.

\begin{figure}[t]
\centering
\begin{minted}[frame=single, fontsize=\footnotesize]{haskell}
unlabel  :: Label l => Labeled l a -> Enclave l p a
unlabelP :: PrivDesc l p => Priv p -> Labeled l a -> Enclave l p a
label    :: (Label l, Binary l, Binary a) 
         => l -> a -> Enclave l p (Labeled l a)
labelP   :: (PrivDesc l p, Binary l, Binary a)
         => Priv p -> l -> a -> Enclave l p (Labeled l a)
taint    :: Label l => l -> Enclave l p ()
taintP   :: PrivDesc l p => Priv p -> l -> Enclave l p ()
\end{minted}
\caption{Core \hasteeplus APIs for Information Flow Control}
\label{fig:difcapi}
\end{figure}

The operations shown above dynamically compute the $\sqsubseteq$ and $\sqsubseteq_P$ relation to determine allowed information flows. 
The \code{PrivDesc} typeclass permits delegating privileges akin to the \textit{acts for} relation described in the Myers-Liskov labeling model \cite{DBLP:journals/tosem/MyersL00}.
The type \code{Labeled l a} exists to allow labeling values to labels other than $L_{cur}$.
Labeled data can be used to indicate data ownership and hence we provide additional APIs for the \code{Client} monad to \code{label}, serialise and \code{unlabel} data, inspired by \textit{labeled communication} in the COWL system \cite{DBLP:conf/osdi/StefanYMRHKM14}.

An implementation challenge arises when integrating the dynamic IFC mechanism with our partitioning tactic, specifically within the \code{inEnclave :: Label l => LIOState l p -> a -> App (Secure a)} function. This function is used to mark a function as trusted and move it into the TEE.
The polymorphic type \code{a} encodes any general function of the form \code{a$_1$ -> a$_2$ -> ... a$_n$ -> Enclave l b}. 
However, the type-checker is unable to unify the \code{l} in \code{LIOState l p} and the \code{l} in \code{Enclave l b}. 
Due to the dynamic nature of our IFC mechanism, a user can mistakenly supply a different label type at runtime, preventing the type-checker from producing a witness.
Accordingly, we use the \code{Data.Dynamic} module of GHC to \textit{dynamically type} the \code{LIOState l p} term. Thus, before evaluating the LIO computation, our evaluator dynamically checks for matching types and, on success, executes the monadic computation.
A notable aspect is that both the program partitioning and IFC enforcement are \textit{implemented as a Haskell library}, which allows us to use these features of the language.

\section{Case Study: A Confidential Data-Analytics Pattern}\label{dcr}

We present a case study in privacy-preserving data analytics, illustrating how a group of mutually distrusting parties can perform analytics without revealing their data to each other. 
We use the core features of \hasteeplus that we have discussed so far - tierless programming, remote attestation and dynamic IFC with privileges for declassification.

\begin{figure}
         \centering
         \fbox{\includegraphics[scale=0.28]{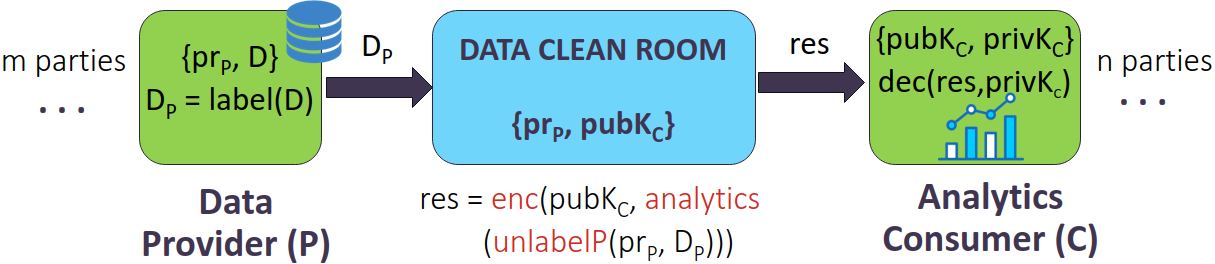}}
         \caption{A \textit{Data Clean Room (DCR)} pattern with $m$ data providers ($P$) and $n$ analytics consumers ($C$). $P$ labels its data as $D_P$ and sends it to the DCR, which loads $C$'s public key $pubK_C$ as well as \textit{privilege} $pr_P$ using a closure. The functions $enc$ and $dec$ handle encryption and decryption, and \textit{analytics} refers to any general query.}
         \label{fig:dcr}
\end{figure}

Fig.~\ref{fig:dcr} shows the overall confidential analytics setup.
The analytics is carried out in a \textit{data clean room (DCR)} - a TEE hosted on a public cloud that aggregates data from multiple parties without revealing the actual data.
In Fig.~\ref{fig:dcr}, the notation $\{a, b, ...\}$ denotes the state (in-memory and persistent) of that party.
The figure shows two distinct sets of participants - the \textit{Data Providers (P)} and the \textit{Analytics Consumers (C)}.
The setup does not limit the number of participants, allowing $m$ data providers and $n$ analytics consumers, where $m, n \in \mathbb{N}$.
Additionally, there are \textit{no restrictions requiring the data providers and the analytics consumer to be different}, and hence in some cases $P = C$.

In this setup, the data providers \textit{label} their data before sending it to the DCR.
The DCR is loaded with analytics queries from $C$ after $P$ reviews both the schema and the specific queries requested by $C$.
The DCR is provisioned with $C$'s public key, limiting access to the computed analytics solely to $C$.
We notably use Haskell's partial application to load the privilege $pr_P$, enabling data unlabeling while restricting the locations of the \code{unlabelP} operation.

Consider a synthetic data analytics example with two data providers $P_1$ and $P_2$ that both store confidential data regarding COVID strains and the corresponding age of patients.
An analytics consumer $C_1$ wishes to aggregate this data and derive the correlation between mean age and the respective COVID strain.
We add a constraint that the analytics should only aggregate COVID strains that are common to both $P_1$ and $P_2$ (\textit{private set intersection} \cite{DBLP:journals/csr/EscaleraAL23}).

The DCR exposes two API calls for communication. 
The first, \code{datasend}, accepts a \code{DCLabeled Row} as an argument and is used by $P_1$ and $P_2$ to send a row labeled with their respective DC Label.
The DCR stores the in-memory data in \hasteeplus's mutable reference type \code{DCRef a}. 
Reading and writing occur via \code{readRef} and \code{writeRef}, raising the context's label accordingly.
We omit the body of \code{datasend} for brevity. 
The full Haskell program is publicly available \cite{hasteedcr}.

Listing~\ref{lst:dcrquery} (lines~1-7) shows the second interface to the DCR, \code{runQuery}, used by $C_1$ to run the analytics query.
A notable operation happens in line~5 where the \code{unLabelFunc} function is applied to each row of the database, labeled with either $P_1$'s or $P_2$'s DC label.
\code{unlabelFunc} inspects the label and accordingly uses the correct privilege to declassify the data. 
Note, if during the whole computation, a row's label gets tainted by both $P_1$ and $P_2$, the \code{unlabel} function (not \code{unlabelP}, see line~13) is invoked, floating the context high enough that writes to public channels are no longer possible.

\begin{listing}[t]
\begin{minted}[fontsize=\footnotesize, linenos]{Haskell}
runQuery :: EnclaveDC (DCRef DB) -> PublicKey 
         -> Priv CNF -> Priv CNF -> EnclaveDC ResultEncrypted
runQuery enc_ref_db pubKC1 priv1 priv2  = do
  labeled_rows <- join $ readRef <$> enc_ref_db
  rows         <- mapM (unlabelFunc priv1 priv2) labeled_rows
  res_enc      <- encrypt_TCB pubKC1 (toStrict $ encode $ query rows)
  return res_enc -- encryption error-handling elided

unlabelFunc p1 p2 lrow =
  case extractOrgName (labelOf lrow) of
    "P1" -> unlabelP p1 lrow
    "P2" -> unlabelP p2 lrow
    _      -> unlabel lrow -- label will float up

data API = API { datasend :: Secure (DCLabeled Row -> EnclaveDC ())
               , runQ     :: Secure (EnclaveDC ResultEncrypted }

app = do db     <- liftNewRef dcPublic database
         sfunc  <- inEnclave dcDefaultState $ sendData db
         pubKC1 <- liftIO $ read <$> readFile "ssl/public.key"
         p1Priv <- liftIO $ privInit (toCNF p1)
         p2Priv <- liftIO $ privInit (toCNF p2)
         qfunc  <- inEnclave dcDefaultState $ runQuery db pubKC1 p1Priv p2Priv
         let api = API sfunc qfunc
         runClient (client1 api)
         runClient (client2 api)
         runClient (client3 api)
\end{minted} 
\caption{\code{runQuery} unlabels the data, runs the query and encrypts the result; The \code{app :: App Done} computation captures the three clients and the enclave}
\label{lst:dcrquery}
\end{listing}

In the absence of privileges, the \code{EnclaveDC} monad obeys general non-interference \cite{DBLP:conf/sp/GoguenM82a}. 
Hence, privileges, which allow \textit{declassification} and \textit{endorsement}, must be handed out with caution and used in limited places.
In \code{app}, the privileges are created (lines~21, 22) and are partially applied to the \code{runQuery} function (line~23).
As a result, the enclave loads a partially applied closure, \code{runQuery db pubKC1 p1Priv p2Priv}, and it is limited to using the privileges solely within \code{runQuery} and its \textit{callees}. 
An interesting future work would be using Haskell's linear type support \cite{DBLP:journals/pacmpl/BernardyBNJS18} to limit the copying of privileges and make them \textit{unforgeable}.

The \code{app} function demonstrates the overall \textit{tierless} nature of our DSL.
It describes three clients and the enclave as a single program without specifying complex data copying protocols or involving multi-project hierarchies. 
We elide the body of the clients $P_1$ and $P_2$, involving data retrieval from their databases, labeling, and sending it to DCR, while $C_1$  calls \code{runQuery} and decrypts the result. 
We also omit the \code{query} function's implementation, responsible for executing private set intersection and returning results in a structured format.
%
% The interested reader is directed to Appendix~\ref{apdx:dcr} for the full program.
The interested reader can find the entire program hosted publicly \cite{hasteedcr}.
In-transit security, enclave-integrity and client-integrity checks are implicitly enforced on all communication through \hasteeplus's remote attestation infrastructure.
% ERASE
% last line of this section and the first line of next similar

% For illustration purposes we show a simplified data processing but Haskell binding exist to powerful machine learning libraries such as TensorFlow https://hackage.haskell.org/package/tensorflow and can be used in such scenarios

\subsection{Security Analysis} \label{discsec}

\textbf{Privacy Protection.} The data clean room ensures privacy through - (1) \textit{runtime security}, provided by the TEE's isolation of trusted code and data; (2) \textit{in-transit security}, ensured by the RA-TLS protocol; (3) \textit{enclave integrity}, established through remote attestation; (4) \textit{client integrity}, provided with digital signatures checked by a monitor (Section~\ref{attestation}); and (5) \textit{information flow control}, implemented using a mix of IFC mechanisms and controlled privilege delegation.

\textbf{Why is the result encrypted if \hasteeplus's monitor already does client-integrity checks?} This is necessary due to two distinct attacker models: \textit{open-world attacks} and \textit{closed-world attacks}. 
The digital signature verification in the monitor protects against open-world attacks, where an unknown malicious attacker outside our described system attempts to communicate with the DCR. 
On the other hand, in a closed-world attack, one of the participating entities, say $P_1$, may maliciously query the DCR for analytics, even though $P_1$ is intended to be merely a data provider. 
Although the monitor will allow this communication, the encryption will protect the data privacy.

\textbf{Declassification Dimensions \cite{DBLP:journals/jcs/SabelfeldS09}} Classifying the DCR along the four dimensions - \textbf{Who}: Integrity checks in \hasteeplus, RA-TLS and data labeling constrain the \textit{who} dimension, allowing the DCR alone to declassify the data; \textbf{What}: The combination of the Haskell type system and dynamic privileges aim to allow declassification only for the analytics query; \textbf{Where}: The partial-application-based privilege loading was done to restrict this dimension, ensuring that only \code{runQuery} and its subsequent \textit{callees} can declassify; \textbf{When} This is currently not captured but it is fairly straightforward to implement a \textit{relative declassification} \cite{DBLP:journals/jcs/SabelfeldS09} policy where the analytics is released only after a certain set of data uploads succeeds, especially using the tierless nature of our DSL.

While \hasteeplus's privacy protection mechanism provides useful guardrails against information leaks, we emphasize the importance of \textit{auditing} the trusted code by all concerned parties to ensure the privileges are not misused.
% Sabelfeld and Sands - Dimensions and principles of declassification
% In here or the case study section?
% What - Analytics Data is Released
% Who  - RA-TLS + Client Integrity check + Enclave that loads the closure with the privilege
% Where - The part of the code with \code{unlabelP} Loading the closure further narrows where \code{unlabelP} is called
% When  - Not captured. Relative - declassify analytics only after org1 has sent data

% One alternative is using homomorphic encryption to work with encrypted data but the performance penalty is immense. Labeling provides an alternative where we hand over data to a \textbf{trusted code base} but to prevent human error and accidental leaks, track the label in a monad

% Simple extension - Use CLIO like labeled key-value store for the database as well

% Replay attacks - probably taken care of by TLS

% List attacks possible in the Remote Attestation space
% Dynamic IFC for dynamic attackers
% Termination and timing channels in LIO \cite{DBLP:conf/icfp/StefanRBLMM12}

% The Binary constraint still acts as a protection

\section{Performance Evaluations}
Here, we present performance microbenchmarks that allow for quantifying the overheads associated with various features in \hasteeplus.
For benchmarking, we use the password checker example from Listing~\ref{lst:pwdchecker}.
We evaluate three particular sources of overheads - (1) \textit{dynamic checks for information flow control}, (2) \textit{remote attestation} and (3) \textit{client-integrity checks} performed by the monitoring module.

\begin{figure}[t]
         \includegraphics[scale=0.55]{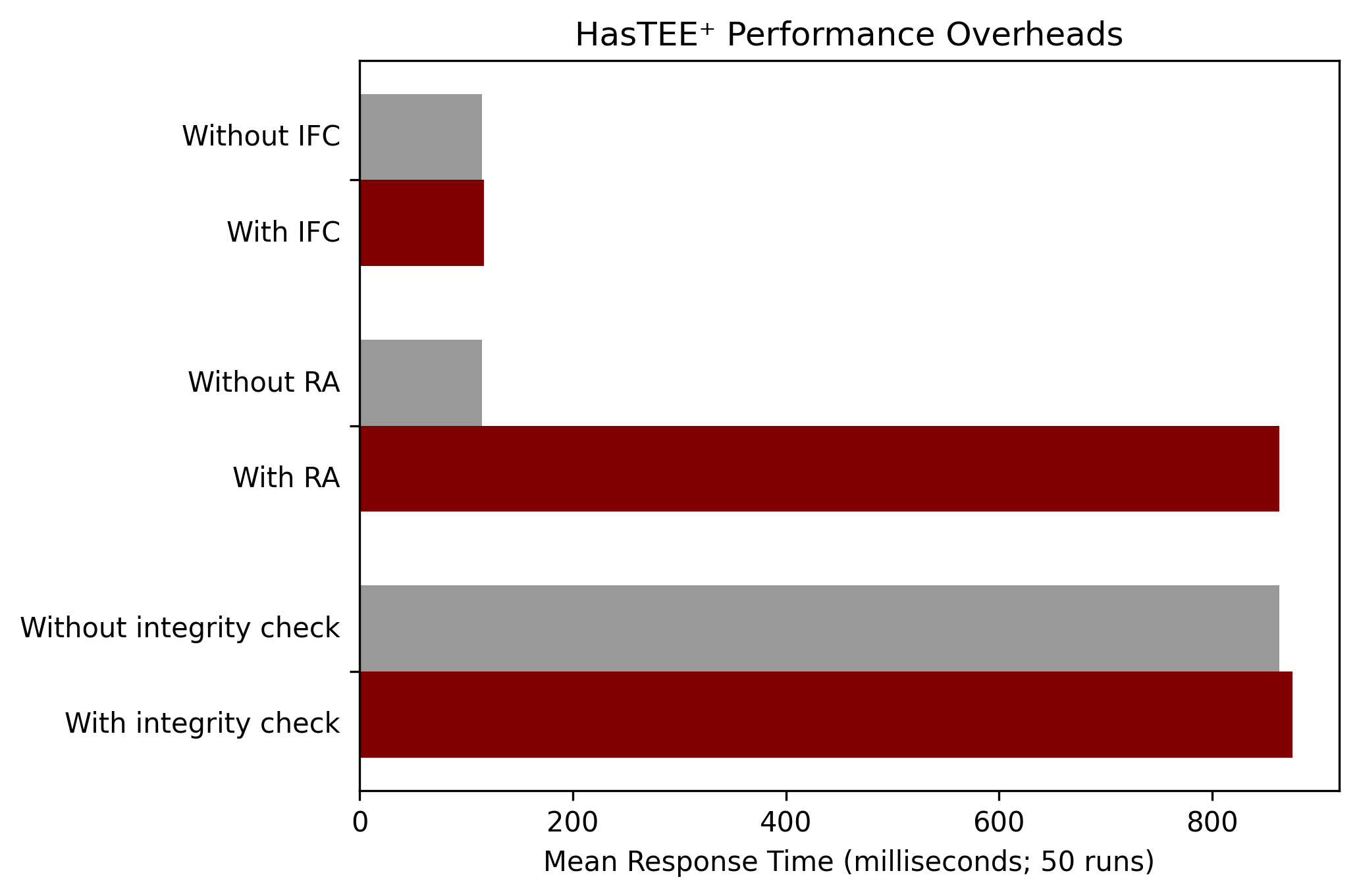}
         \caption{Performance Overheads in \hasteeplus. The X-axis represents the mean response time for a query, both with and without the desired feature enabled, measured in milliseconds. The response time is averaged over 50 runs for each measure.}
         \label{fig:perf}
\end{figure}

We plot the overheads associated with each feature in Fig.~\ref{fig:perf}.
The X-axis captures the mean response-time in executing variants of the operation \code{gatewayRA ((checkpwd api) <@> userInput)} (line~14) from Listing~\ref{lst:pwdchecker}.
A \textit{gateway} call involves serialising the function arguments, making a remote procedure call to the enclave, executing the enclave computation, and then deserialising the result to the client.
Measurement were conducted on an Azure Standard DC1s v2 (1 vcpu, 4 GiB memory) SGX machine, using the Intel SGX SDK for Linux.

\textbf{Dynamic IFC Checks Overhead.} In Fig.~\ref{fig:perf}, the first group measures the overhead due to the runtime checks for IFC. 
The difference arises from extra conditional statements checking legal data flow policies.
For Listing~\ref{lst:pwdchecker}, the overhead is 2 milliseconds when compared to the non-IFC version.
% The overheads are minimal for Listing~\ref{lst:pwdchecker} with a difference of only 2 milliseconds compared to the non-IFC version.
%
However, more complex applications (like Section~\ref{dcr}) might incur slightly higher overheads.
% However, applications with more complex flows (like Section~\ref{dcr}) might incur slightly higher overheads.

\textbf{Remote Attestation (RA) Overhead.} The second group shows the RA overhead, demonstrating a considerable jump in response time with RA enabled.
Much of the latency is attributed to the underlying \textit{RA-TLS} protocol, implementing TLS version 1.2.
In contrast, the non-RA baseline employs plain TCP for communication, using Linux's \code{send/recv} to enhance communication speed.
The RA version's mean latency is 863 milliseconds, improvable by establishing a secure channel instead of initiating the entire handshake protocol each time.

\textbf{Integrity-check Overhead.} The client-integrity check is built on top of the \hasteeplus RA infrastructure.
As a result, for the baseline we use RA measurements from the second group and incorporate integrity checks on top of RA. 
The overhead on top of RA is minimal, in the order of 15 milliseconds.
% , as seen in Fig.~\ref{fig:perf}.
%
% Currently, the integrity check is opted in with a \code{-fintegrity-check} flag.

% \textbf{Developer Effort.}
%Evaluation point - in terms of reducing developer effort; LOC in a remote attestation app in C vs remote attestation with HasTEE; time taken by the authors

\textbf{Discussion.} 
The measurements in Fig~\ref{fig:perf} show that each \hasteeplus feature incurs maximum overheads in the order of hundreds of milliseconds.
The significant response time increase for RA is mainly due to the complex TLS handshake involving multiple hops and communication with the Intel Attestation Service.
Given the security-critical nature of confidential computing and considering slowdowns due to general network latency, we posit that \textit{\hasteeplus's overheads are acceptable, making it a practical choice for security-critical applications.}

\section{Related Work}

We have already discussed projects closely related to \hasteeplus, including HasTEE \cite{DBLP:conf/haskell/SarkarKRC23}, GoTEE \cite{DBLP:conf/usenix/GhosnLB19}, $J_E$ \cite{DBLP:conf/csfw/OakABS21}, etc., in Sections~\ref{sec:intro} and \ref{sec:threat}.
Here, we highlight additional related work that is relevant to the broader contributions of \hasteeplus.

\textbf{Tierless Programming for Enclaves} To the best of our knowledge, \hasteeplus is one of the first practical programming frameworks to introduce the notion of \textit{tierless} programming for confidential computing applications.
Weisenburger et al. \cite{DBLP:journals/csur/WeisenburgerWS20} provide a survey of general multi-tier programming approaches.
Among the surveyed approaches, the \hasteeplus DSL draws inspiration from the \textit{Haste framework} \cite{DBLP:conf/haskell/EkbladC14} and functional choreographic programming \cite{DBLP:journals/pacmpl/ShenKK23}.

\textbf{Remote Attestation Infrastructure}. \hasteeplus is built on top of the Intel RA-TLS protocol \cite{knauth2018integrating} for \textit{binary attestation}. 
In contrast, GuaranTEE \cite{DBLP:conf/IEEEcloud/MorbitzerKZ23} proposes a \textit{control-flow attestation} technique based on two enclaves, which can be adapted quite naturally to the \hasteeplus RA infrastructure.

\textbf{Information Flow Control for TEEs.} Gollamudi et al. proposed the first use of IFC to protect against low-level attackers in TEEs with the IMP$_E$ calculus \cite{DBLP:conf/oopsla/GollamudiC16}, followed by a more general security calculus, DFLATE \cite{DBLP:conf/csfw/GollamudiCA19}, for distributed TEE applications. 
In contrast to their work, \hasteeplus does not require language-level modifications or type-system extensions. 
Instead, it conveniently enforces \textit{IFC as a library} in an existing programming language.
At the OS level, Deluminator \cite{DBLP:conf/raid/TarkhaniM23} offers OS abstractions and userspace APIs for trace-based tracking of IFC violations in compartmentalized hardware, such as TEEs. Note that Deluminator is a reporting tool and not an \textit{enforcement} mechanism, in contrast to \hasteeplus.
Another application of IFC to TEEs is Moat \cite{DBLP:conf/ccs/SinhaRSV15}, which formally verifies the confidentiality of enclave programs by proving the non-interference property \cite{DBLP:conf/sp/GoguenM82a}.

%% ERASE for SPACE
\textbf{Confidential Data Analytics.} Referring to our confidential analytics pattern in Section~\ref{dcr}, another proposed design pattern is Privacy Preserving Federated Learning \cite{DBLP:conf/mobisys/MoHKMPK21}, tailored to machine learning attacker models.
We believe such threat models can be naturally integrated with our proposed design pattern.

% Closest relative of \hasteeplus - $J_E$ \cite{DBLP:conf/csfw/OakABS21}; lacks remote attestation; no tierless DSL; $J_E$ is in spirit closer to HasTEE, GoTEE; etc its an SGX SDK with IFC

% $J_E$ is best suited to adopt the remote attestation infrastructure proposed here; The IFC mechanism they use is slightly different and would be interesting to see if they can adopt \hasteeplus's style of I/O. Can their partitioning represent multi-client-server? What kind of data flow analysis do they do to partition programs?

\vspace{-1mm}
\section{Conclusion}
We introduced \hasteeplus, a \textit{tierless} confidential computing DSL that enforces dynamic information flow control, along with strong client-integrity and enclave-integrity checks. 
We also proposed a general confidential analytics pattern, expressed as a single program in \hasteeplus. Additionally, we presented performance evaluations that demonstrate acceptable overheads.
Our evaluations, while conducted on Intel SGX, illustrate \hasteeplus's general applicability to ARM TrustZone, AMD SEV, and Intel TDX machines. 
Furthermore, our library-based partitioning and IFC approach is extendable to other programming languages.

\bibliographystyle{splncs04}
\bibliography{mybiblio}

\end{document}